\documentclass[oldversion]{aa} 

\usepackage{graphicx}
\usepackage{txfonts}
\usepackage{natbib}
\bibpunct{(}{)}{;}{a}{}{,} 
\begin{document}
\title{CEN\,34 -- High-Mass YSO in M\,17 or Background Post-AGB Star ?
\thanks{Based on observations performed with the
    ESO \textsl{Very large Telescope} on Cerro Paranal, Chile, as part
    of the X-shooter Science Verification program 60.A-9402(A).}}

\subtitle{}

\author{Zhiwei Chen\inst{1,2,3}
\and Dieter N{\"u}rnberger~\inst{2} \and  Rolf Chini\inst{2,4} \and Yao Liu\inst{1} \and Min Fang\inst{1} \and Zhibo Jiang\inst{1}}

\institute{Purple Mountain Observatory, Chinese Academy of Sciences, 2 West Beijing Road, 210008 Nanjing, China \\
\email{zwchen@pmo.ac.cn}
\and Astronomisches Institut, Ruhr--Universit{\"a}t Bochum, Universit{\"a}tsstrasse 150, 44801 Bochum, Germany \and University of Chinese Academy of Sciences, 100039 Beijing, China \and Instituto de Astronom{\'i}a, Universidad Cat{\'o}lica del Norte, Avenida Angamos 0610, Casilla 1280 Antofagasta, Chile\\
  }

\date{Received 15 April 2013 / Accepted 11 July 2013}


\abstract{We investigate the proposed high-mass young stellar object\,(YSO) candidate CEN\,34, thought to be associated with the star forming region M\,17. Its optical to near-infrared ($550 - 2500$\,nm) spectrum reveals several photospheric absorption features, such as H$\alpha$, the \ion{Ca}{ii} triplet and the CO bandheads but lacks any emission lines. The spectral features in the range $8375 - 8770$\,{\AA}, showing the highest S/N ratio, are used to constrain an effective temperature $T_\mathrm{eff} = 5250 \pm 250\,\mathrm{K}$ (early-/mid-G) and a $\log g = 2.0 \pm 0.3$ (supergiant). The spectral energy distribution displays a faint infrared excess that resembles the SED of a high-mass YSO or an evolved star of intermediate mass. Moreover, the observed temperature and surface gravity are identical for high-mass YSOs and evolved stars. The radial velocity relative to LSR~($V_{LSR}$) of CEN\,34 as obtained from various photospheric lines is of the order of $-60$\,km/s and thus distinct from the $+25$\,km/s found for several OB stars in the cluster and for the associated molecular cloud. The SED modeling yields $\sim10^{-4}M_{\sun}$ of circumstellar material which contributes only a tiny fraction to the total visual extinction~(11 mag). The distance of CEN\,34 is between 2.0~kpc and 4.5~kpc. In the case of a YSO, a dynamical ejection process is proposed to explain the $V_{LSR}$ difference between CEN\,34 and  M\,17. Additionally, to match the temperature and luminosity, we speculate that CEN\,34 had accumulated the bulk of its mass with accretion rate $> 4\times10^{-3}M_{\sun}/$yr in a very short time span~($\sim10^3$~yrs), and currently undergoes a phase of gravitational contraction without any further mass gain. However, all the aforementioned characteristics of CEN\,34 are compatible with an evolved star of $5 - 7\,M_{\sun}$ and an age of $50 - 100$\,Myrs, most likely a background post-AGB star with a distance between 2.0 kpc and 4.5 kpc. We consider the latter classification as the more likely interpretation. Further discrimination between the two possible scenarios should come from the more strict confinement of CEN\,34's distance.

  }

\maketitle

\section{Introduction}

\begin{figure*}
 \centering
 \includegraphics[width=0.7\textwidth]{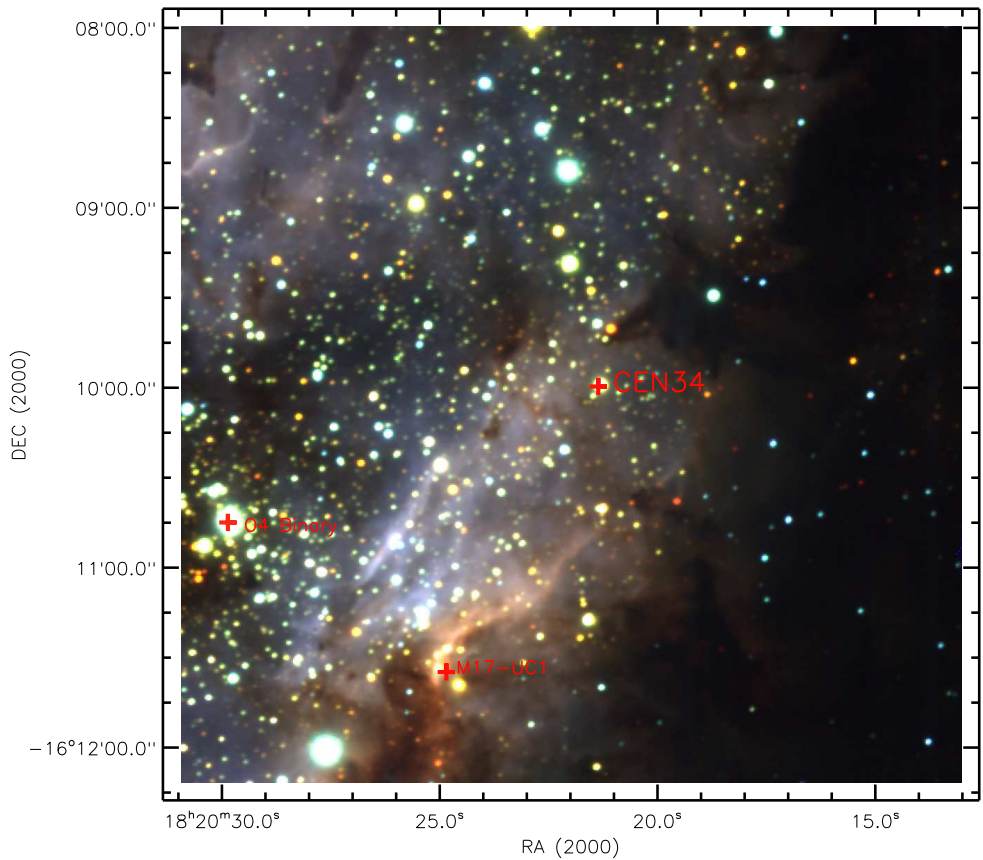}
 \caption{The projection of CEN\,34~(red cross) on the sky overlaid on the $JHK_s$ pseudo-color image of M\,17 taken by the IRSF~(\,{\bf I}nf{\bf R}ared {\bf S}urvey {\bf F}acility) telescope. The ionizing sources of M\,17--the O4-type binary, and the hypercompact \ion{H}{ii} region--M\,17-UC1 are marked also.}

  \label{Fig:Fd}%

\end{figure*}

Before reaching the main sequence, young stars accumulate the bulk of their mass during the so-called pre-main sequence (PMS) evolutionary phase. This phase can last as long as a few Myr for low and intermediate mass stars ($<$ 8\,$M_{\sun}$), while it can be as short as $\sim10^4-10^5$\,yr for high-mass stars ($>$ 8\,$M_{\sun}$) \citep{2007ARA&A..45..481Z}. During the majority of their PMS phase young stars are surrounded and obscured by significant amounts of molecular gas and dust; the younger the forming stars the more deeply embedded they are in their natal environments and, thus, the more difficult is their detection and characterization through observations at optical and near-IR wavelength \citep[e.g.,][]{2007ApJ...656L..81N}. The detection of high-mass YSOs is rare, because these objects are deeply embedded, but equally hampering is the extremely short PMS evolutionary phase, which makes it statistically more difficult to find them. Furthermore, the search for and the identification of high-mass YSOs is hampered by confusion with much more evolved objects, showing similar or even identical observational signatures. For instance, if infrared~(IR) excess is considered as sole criteria for source selection, high-mass YSO candidates and post-asymptotic giant branch (post-AGB) stars might be easily mixed up, although they are totally distinct in evolutionary phase. In both cases the observed IR excess originates from absorption of stellar photons by circumstellar dust and subsequent re-emission as thermal radiation at longer wavelengths. Alternatively, both the forming high-mass YSOs and the evolved stars have larger stellar radii and correspondingly cooler effective temperatures than the main sequence stars. This also prevents distinguishing high-mass YSOs from evolved stars solely from a spectroscopic point-of-view.

As one of the brightest \ion{H}{ii} regions in the Galaxy, M\,17 shows evidence for multiple epochs of massive star formation~\citep{2002ApJ...577..245J,2009ApJ...696.1278P,2012PASJ...64...110C}. It features two prominent bars, the northern \ion{H}{ii} bar~(N-bar) and the southwestern \ion{H}{ii} bar~(S-bar), which are ionized by a couple of OB-type stars located in the center of the \ion{H}{ii} region \citep{2012PASJ...64...110C}. A number of intermediate and high-mass YSOs have been discovered in the S-bar~\citep{2004Natur.429..155C,2007A&A...465..931N,2007ApJ...656L..81N,2011A&A...536L...1O}. In addition, the S-bar harbors several high-mass YSO candidates, which display rising SEDs towards IR wavelengths but still await confirmation of their preliminary classification \citep{2001A&A...377..273N}. Among these high-mass YSO candidates, CEN\,34 (also known as B\,358) has its own confusing peculiarities. It is very bright in the near IR ($K(\mathrm{2MASS})=7.88$\,mag) and rather faint at optical wavelengths~(B\,$\sim$\,22.8\,mag, V\,$\sim$\,18.8\,mag) indicative of high extinction. Its $K$-band spectrum shows the CO bandheads longwards of 2.29\,$\mu$m in absorption \citep{1997ApJ...489..698H,2006A&A...457L..29H}, reminiscent of late-type Class II/III YSOs~\citep{1996A&A...306..427C,2006A&A...457L..29H}. However, the same feature is typical for late-type stars as well as for evolved red stars~\citep[e.g.,][]{1997ApJS..111..445W}. Therefore, the true nature of CEN\,34 is unclear.

To constrain effective temperature and luminosity class of CEN\,34, which are crucial to better understand its nature, we have performed new spectroscopic observations with high sensitivity and large wavelength coverage (from the Optical to the Near-IR), using the medium-resolution spectrograph X-Shooter on ESO's \textsl{Very Large Telescope}~(\textsl{VLT}). In this paper we report on our findings with respect to CEN\,34's spectral classification, its radial velocity in comparison to that of other cluster members, its extinction and its most likely distance.

\begin{figure*}
   \centering
   \includegraphics[width=0.4\textwidth,angle=90]{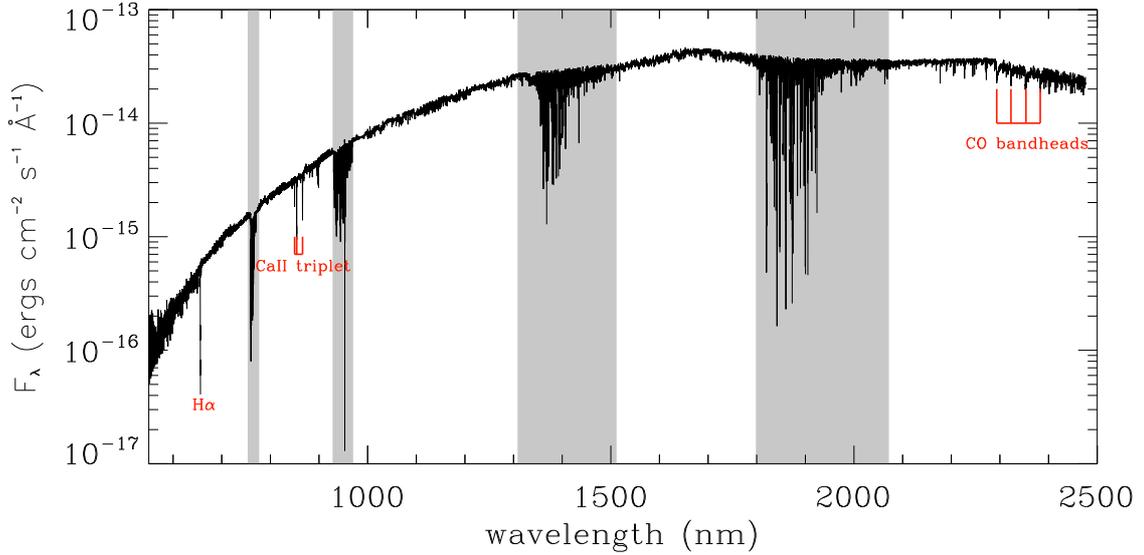}
   \caption{X-Shooter spectrum of CEN\,34, covering the visible and near-IR wavelength range. Prominent photospheric absorption lines are marked. Grey shaded regions mark the major telluric absorption features.}
   \label{Fig:Xs}%
 \end{figure*}

\section{Observations and Data Reduction\label{Sec:Od}}

The \textsl{VLT}/X-Shooter observations of CEN\,34~($18^{\mathrm{h}}20^{\mathrm{m}}21\fs36$,
~$-16\degr09\arcmin59\farcs6$~\textrm{J2000}) were performed on August 12 2009, during the first science verification run of X-Shooter, which is the first of the second generation instruments at VLT~\citep{2006SPIE.6269E..98D,2011A&A...536A.105V}. The average optical seeing and airmass during the observations are $0\farcs59$ and 1.376, respectively. All observations were acquired by nodding the star on the slit using a ABBA sequence, allowing for proper sky subtraction and bad pixels removal. On every nodding pointing, the integration time settings are DIT$=$685~s and NDIT$=$1 for the UVB arm~(300--600~nm), DIT$=$285~s and NDIT$=$2 for the VIS arm~(550--1000~nm),  DIT$=$11~s and NDIT$=$3 for the NIR arm~(1000--2500~nm). In total, the on-source integration time is 45 minutes for UVB arm, 38 minutes for VIS arm, and 2.2 minutes for NIR arm. The data taken in the UVB arm were not satisfying because the faintness of CEN\,34 provided an insufficient S/N. For the VIS arm and the NIR arm a $0\farcs9$ slit width was used to guarantee $R$\,$=$\,8800 ($\Delta\lambda=0.09$\,nm) and $R$\,$=$\,5600 ($\Delta\lambda=0.3$~nm), respectively. The S/N ratio is $\sim 30$ in the VIS arm and $\sim 20$ in the NIR arm. The standard procedures of data reduction were applied using the X-Shooter pipeline version 1.3.7~\citep{2011AN....332..227G}, including bias, dark and flat-field correction, wavelength calibration, order tracing and flux calibration. Regarding the wavelength calibration, the transformation from pixel space to ($\lambda$, s) space is performed through the analysis of a 9-pinhole Th-Ar lamp frame with a list of lines and a table containing the expected positions of the lines in the frame. The accuracy of wavelength calibration using the pipeline is better than $\sim2$~km/s over the entire wavelength range of X-Shooter. The standard star \object{EG\,274}~(a white dwarf) was used to calibrate the fluxes for the spectra of the three individual arms. Meanwhile, the telluric absorption correction was conducted for the spectrum of the NIR arm by using the telluric standard stars \object{HD\,100858}.

\section{Results\label{Sec:Re}}
The relative location of CEN\,34~(red cross) with respect to the \ion{H}{ii} region is shown in the finding chart~(see Fig.~\ref{Fig:Fd}). The most massive stars, the O\,4-type binary~\citep{2008ApJ...686..310H}, are located at the left edge of the finding chart, meanwhile the giant molecular cloud M\,17-SW is located at the right edge.

The flux calibrated spectrum of CEN\,34 from the visual to the near-IR is shown in Fig.~\ref{Fig:Xs}. The optical and near-IR parts of the spectrum are quite noisy due to the faintness at optical wavelengths and the short exposure time at near-IR wavelengths, respectively. Only absorption lines are detected, such as H$\alpha$, \ion{Ca}{ii} triplet, and CO bandheads in $K$-band. The analysis based on the \ion{Ca}{ii} triplet and CO bandheads are presented in the following subsections.

\subsection{Spectral Type and Luminosity Class\label{S:Spt}}

Our spectral classification of CEN\,34 relies on the wavelength range 8375--8770\,{\AA} which avoids strong telluric absorption. It also offers a rich combination of lines relevant for astrophysical diagnostics and spectral classification \citep{2009ssc..book.....G}. First, neutral Fe and Ti lines appear from late-F stars onwards, and grow gradually in intensity with decreasing temperature until late-K stars. Second, the Paschen series~(e.g., Paschen 14 line 8598\,{\AA}, hereafter P14) disappear for main sequence stars later than G0, but they can still be seen for mid-G supergiants due to the luminosity effect~\citep{1999A&AS..137..521M}. Third, neutral Ti and Fe lines, the \ion{Ca}{ii} triplet~(8498~{\AA}, 8542~{\AA}, 8662~{\AA}) as well as the lines of the Paschen series can provide luminosity discrimination by comparing their absolute strengths.

\begin{figure}
   \centering
   \includegraphics[width=0.3\textwidth,angle=90]{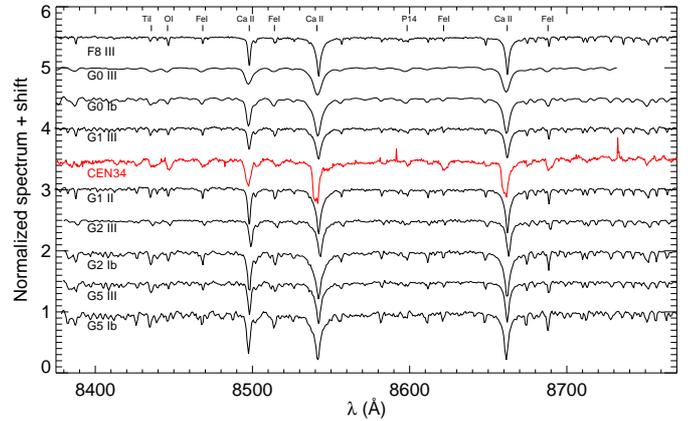}
   \caption{The X-Shooter spectrum of CEN\,34~(emphasized by red color) in the wavelength region 8375--8770~{\AA}. The two features apparently seen in emission in the CEN\,34 spectrum at wavelengths $\sim8590$\,{\AA} and $\sim8730$\,{\AA} are artefacts with unnatural line widths. For comparison, the spectra of several reference stars are displayed; they are adopted from the literature as follows: F8~III~(\object{HIP\,74975}), G1~III~(\object{HIP\,117503}) and G1~II~(\object{HIP\,47908}) from \citet{2004ApJS..152..251V}, G0~Ib~(\object{HIP\,96481}) and G0~III~(\object{HIP\,5454}) from \citet{1995A&AS..112..475A}, G2~Ib~(\object{HIP\,109074}), G2~III~(\object{HIP\,86731}), G5~III~(\object{HIP 100524}), and G5~Ib~(\object{HIP 107348}) from \citet{1997A&AS..123....5C}. All reference spectra have a resolution around 1~{\AA}, very close to that of our CEN\,34 spectrum. All spectra are normalized and are displayed by applying a corresponding shift along the Y-axis for better visual comparison.}
   \label{Fig:Sp}%
 \end{figure}

\begin{table}
  \caption{Line Parameters}
  \label{T:LINES}
  \centering
  \begin{tabular}{c c c }
    \hline\hline
    $\lambda_{obs}~(\lambda_0)$ & Ion  & Intensity \\
    $[\AA]$ &  & in absorption \\
    \hline
    8437 & \ion{Ti}{i} & 0.10 \\
    8446 & \ion{O}{i}  & 0.15 \\
    8468 & \ion{Fe}{i} & 0.10 \\
    8496.6~(8498.02) & \ion{Ca}{ii} & 0.43 \\
    8515 & \ion{Fe}{i} & 0.09 \\
    8540.3~(8542.09) & \ion{Ca}{ii} & 0.72 \\
    8598 & \ion{H}{i} & 0.09 \\
    8621 & \ion{Fe}{i} & 0.15 \\
    8660.5~(8662.14) & \ion{Ca}{ii} & 0.61 \\
    8689 & \ion{Fe}{i} & 0.18 \\
    \hline
  \end{tabular}
  \tablefoot{The intensities of the lines are normalized line strengths with respect to the continuum. For the \ion{Ca}{ii} triplet the references of rest wavelength in air~\citep[within the bracket;][]{1980wtpa.book.....R} are given for the purpose of calculating the $V_{LSR}$ of CEN\,34. The weaker lines have $S/N$ ratios between 5 and 9 and were not used for velocity determinations; therefore their rest frame wavelengths are not listed.}
\end{table}

Fig.~\ref{Fig:Sp} shows the wavelength range 8375--8770~{\AA} of the X-Shooter spectrum of CEN\,34~(highlighted in red color) in comparison to the spectra of several reference stars of luminosity classes I~(supergiants), II~(bright giants) and III~(normal giants). Prominent absorption lines -- like the \ion{Ca}{ii} triplet, neutral metal lines (\ion{Fe}{i}, \ion{Ti}{i}, and \ion{O}{i}), and the P14 line -- are marked; their derived parameters are listed in Table~\ref{T:LINES}. The appearance of \ion{Fe}{i}, \ion{Ti}{i} and Paschen lines in the spectrum of CEN\,34 indicates a rough temperature class of late-F to mid-G. The absolute strengths of these absorption lines are greater than those for giants, suggesting a luminosity class II or I. In addition, \ion{Ti}{i} and Paschen lines stronger than in the spectrum of the F8\,III reference star can be due to brighter luminosity class (II or I) and/or later temperature class.

\begin{figure*}
  \centering
  \includegraphics[width=0.95\textwidth]{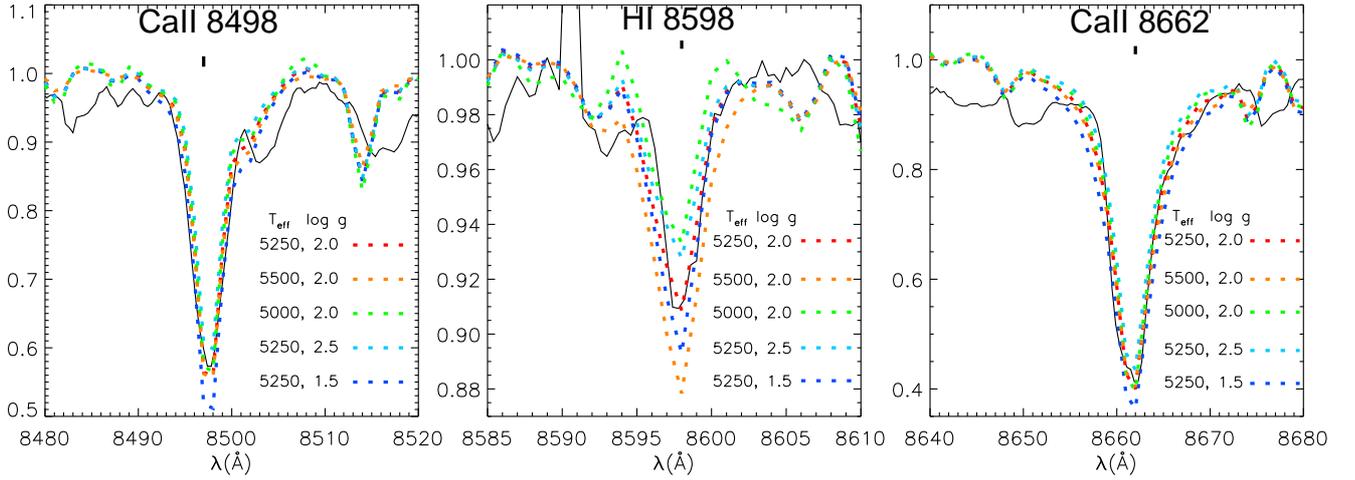}
  \caption{CEN\,34's X-Shooter spectrum (black solid) around the \ion{Ca}{ii}\,8498\,{\AA}, P14, and \ion{Ca}{ii}\,8662\,{\AA} lines. The best fitting grid~(red) as well as other nearby grids are shown as dashed lines. The synthetic spectral library is based on the Kurucz's codes, and has an uniform dispersion of 1~{\AA}/pix~(For more details the reader is referred to \citet{2005A&A...442.1127M}).}
  \label{Fig:Teff}%
\end{figure*}

\subsection{Effective Temperature and Surface Gravity\label{S:Temp}}

Effective temperature and gravity of CEN\,34 are further constrained by comparison of the observed spectrum to the library of synthetic spectra published by \citet{2005A&A...442.1127M} and assuming solar metallicity. The metallicity dependent \ion{Ca}{ii} triplet is used as primary discriminator because of its line strength and best $S/N$ ratio. In addition, to rule out any  metallicity effect, the metallicity independent P14 line is used as secondary discriminator.

We construct a grid of synthetic spectra covering the parameter space $5000~\mathrm{K}\leqslant T_{\mathrm{eff}}\leqslant6000~\mathrm{K},~\Delta T=250~\mathrm{K}$, $0.0\leqslant \log g \leqslant 3.0,~\Delta \log g = 0.5$. On basis of the spectral type of CEN\,34, we suggest a rotation velocity of 10\,km/s~\citep[the typical value for early-G supergiants;][]{2002A&A...395...97D} during the fitting procedure. The best fitting synthetic spectrum should firstly match the \ion{Ca}{ii} triplet, then secondly match the P14 line; fitting results are checked by visual inspection only. Fig.~\ref{Fig:Teff} shows the best fitting grid and other nearby grids. Among the five different grids, we notice that the line strengths of the \ion{Ca}{ii} lines are very sensitive to $\log g$, while are not varying much with $T_\mathrm{eff}$ changing; in contrast, P14 line is sensitive to both $\log g$ and $T_\mathrm{eff}$. Only the grid of $T_\mathrm{eff} = 5250$\,K, $\log g = 2.0$ provides an acceptable match for the two \ion{Ca}{ii} lines as well as the P14 line. Because the $S/N$ ratios of the two \ion{Ca}{ii} lines are much better than that of the P14 line, $\log g$ is better fitted than $T_\mathrm{eff}$. Therefore, the uncertainty of $\log g$ is estimated to be 0.3, smaller than the grid step, but the uncertainty estimate of $T_\mathrm{eff}$ is 250\,K, similar as the grid step. Through this approach, $T_\mathrm{eff}$ and $\log g$ of CEN\,34 are determined as $5250 \pm 250~\mathrm{K}$ and $2.0 \pm 0.3$, respectively.

\subsection{CO\,2-0 Bandhead Absorption\label{S:Co}}

The CO\,2-0 bandhead absorption longwards of 2.29~$\mu$m is typical for late-type stars, whose outer atmospheric layers have the correct temperatures~($\sim 1000 - 3000$\,K) to produce such features. The CO bandhead absorption generally start to appear in the spectra of late-G dwarfs, but their appearance could shift to earlier spectral types for supergiants~\citep[e.g.,][]{1997ApJS..111..445W}. However, low-mass Class II/III YSOs may also exhibit CO bandheads in absorption, because the effective temperature is not high enough to dissociate CO molecules and the circumstellar veiling is less than that of Class I YSOs~\citep{1996A&A...306..427C,2006A&A...457L..29H}.

\begin{figure}
  \centering
  \includegraphics[width=0.45\textwidth]{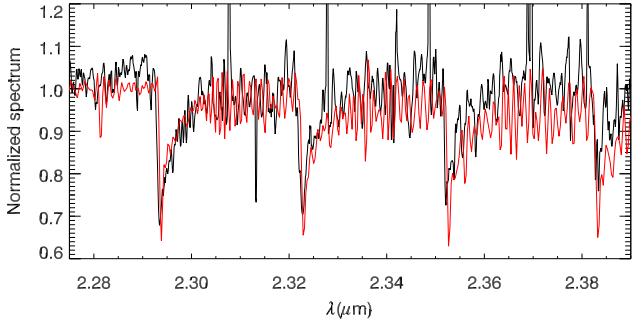}
  \caption{Normalized X-Shooter spectrum of CEN\,34 (black) in the wavelength range 2.28--2.39~$\mu$m, covering the CO\,2-0 bandhead absorption. For comparison, the reference spectrum of the mid-G supergiant \object{HIP\,107348} \citep[SpT G5\,Ib;][]{1997ApJS..111..445W} is shown red. The spectrum of CEN\,34 has been smoothed to the spectral resolution ($R\sim3000$) of the reference spectrum.} 
  \label{Fig:Co}%
\end{figure}

Fig.~\ref{Fig:Co} shows the CO bandhead absorption for CEN\,34 as well as  a reference mid-G supergiant. Since the CO bandhead absorption is strongly related to the $T_\mathrm{eff}$ and to the surface gravity, \citet[][and reference therein]{2012A&A...539A.100G} found a good correlation between the equivalent width of the CO first overtone in absorption~(hereafter $\mathrm{CO}_\mathrm{EW}$) and spectral types for giants and supergiants: $\mathrm{CO}_\mathrm{EW}$ linearly increases with the decreasing $T_\mathrm{eff}$, and is larger for supergiants than for giants of the same $T_\mathrm{eff}$. Therefore the CO bandheads can be used to verify the above spectral type and luminosity class of CEN\,34. Following Fig.~2 in \citet{2012A&A...539A.100G}, the $\mathrm{CO}_\mathrm{EW}\approx11~{\AA}$ of CEN\,34 suggests an early-K giant or mid-G supergiant, the latter agrees well with the previously classified spectral type and luminosity class if the involved uncertainty is considered. 

Based on the spectral classification and synthetic spectral fitting presented above, CEN\,34 has a spectral type of early-/mid-G ($T_\mathrm{eff} = 5250 \pm 250$\,K) and $\log g = 2.0 \pm 0.3$ compatible with a supergiant.

\subsection{Radial Velocity Relative to LSR~($V_{LSR}$)\label{S:Rv}}

Usually, newly formed stars conserve some kinematic imprint of their natal molecular cloud, i.e., stars having originated from the same cloud are expected to display similar radial velocities \citep{1977ApJ...214..747H}. Therefore, $V_{LSR}$ measurements are an excellent tool to investigate membership of stars in star forming regions or stellar clusters.

In the case of M\,17, comparison can be made to the young O- and B-type stars whose membership is well established. For many of them, reliable spectral classification has been derived by \citet{2008ApJ...686..310H}, primarily on the basis of low resolution near-infrared~(NIR) spectroscopy with ISAAC complemented by higher resolution ($R = 6600$) optical-red \textsl{GIRAFFE} spectra covering the wavelength range $8200 - 9400$\,{\AA}. While the spectra of the B-type stars are featureless in the $8000 - 9000$\,{\AA} range, accurate reference $V_{LSR}$ values can be derived for 11 O-type stars with strong \ion{Ar}{I} absorption lines at 8620\,{\AA}. In addition, the X-Shooter spectrum of the B7-type PMS star CEN\,24/B\,275~also serves as $V_{LSR}$ reference~\citep{2011A&A...536L...1O}; its Balmer series yield  $V_{LSR}\sim 24.7$\,km/s). In total, we have calculated the $V_{LSR}$ for 12 well-established OB-type members of M\,17; their mean $V_{LSR}$ is {\bf \boldmath$+29.4$}\,km/s, with a scatter of 7.1\,km/s. Furthermore, the prevalent $V_{LSR}$ of the associated molecular clouds (M17-SW and M17-North) as obtained from radio and millimeter observations~\citep{2003ApJ...590..895W,2011ApJ...733...25X} is {\bf \boldmath$\sim +20$}\,km/s.

\begin{figure}
  \centering
  \includegraphics[width=0.45\textwidth]{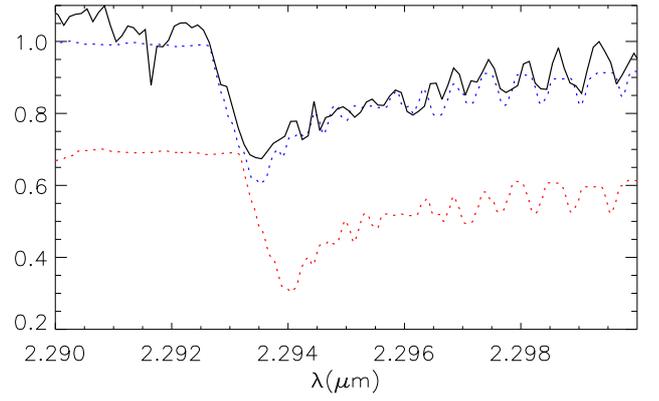}
  \caption{The normalized CO 2-0 first overtone feature of CEN\,34~(black solid) and those of Arcturus~(red and blue dotted lines). The red-dotted presents the rest frame spectrum of Arcturus, while the blue-dotted is the spectrum shifted by the velocity corresponding to the result from the cross-correlation with respect to CEN\,34's spectrum. The rest frame spectrum is shifted by
    arbitrary amounts along the vertical axis for presentation
    purposes.}
  \label{Fig:Rv}%
\end{figure}

To determine the $V_{LSR}$ of CEN\,34 we use the \ion{Ca}{ii} triplet and the CO first overtone in the X-Shooter spectrum. The lines of the triplet are fitted with gaussian absorption profiles to derive their central wavelengths~(see Table~\ref{T:LINES}). Assuming $R = 8800$, the velocity resolution of the X-Shooter VIS arm is around 34\,km/s. Indeed, the central wavelengths fitted by gaussian absorption profiles have errors less than the velocity resolution. By testing different fitting parameters, the wavelength deviations are about 0.2\,{\AA}, corresponding to velocity resolution of 7\,km/s, i.e., 20\% of the velocity resolution. Comparison to the rest line wavelengths in air of the \ion{Ca}{ii} triplet~(see Table~\ref{T:LINES}) then yields a $V_{LSR}$ of {\bf \boldmath$-57 \pm 6$}\,km/s. To obtain an alternative estimate of its $V_{LSR}$ from the CO first overtone, we cross-correlated CEN\,34's spectrum with a high resolution ($R = 100,000$) vacuum rest frame spectrum of Arcturus~(see Fig.~\ref{Fig:Rv}). For more details of this approach the reader is referred to \citet{2003ApJ...599.1139F}. The spectrum of Arcturus was first resampled to match the lower spectral resolution~($R = 5500$) of the X-Shooter NIR arm and then shifted in velocity by {\bf \boldmath$-64$}\,km/s to overlap with the spectrum of CEN\,34. The uncertainties involved in the velocity shift are dominated by the lower spectral resolution of the X-Shooter spectrum and as before assumed to be of the order of 20\% of the velocity resolution, i.e. 11\,km/s. Obviously, the two independent $V_{LSR}$ estimates for CEN\,34 agree reasonably well within the errors. Thus, CEN\,34 has a $V_{LSR}$ which is significantly distinct from those of the OB stars and from those of the associated molecular clouds. This suggests that CEN\,34 is not a member of M\,17 although one could argue that the high-mass stars in M\,17 did conserve their original velocity while stars of lower mass have been kicked out by tidal interactions.

The $V_{LSR}$ of CEN\,34 differs from that of any other molecular cloud within the inner Galaxy along the same line-of-sight~($l = 15\degr$), adopting the Galactic rotation model of \citet{2009ApJ...700..137R}. A spiral arm lying beyond the Outer Arm in the first Galactic quadrant $\sim 23$\,kpc from the Galactic center shows a LSR velocity of $-15$\,km/s at $l = 15\degr$~\citep{2011ApJ...734L..24D}. Therefore, it is not possible to estimate a kinematic distance for CEN\,34 by its $V_{LSR}$.

\begin{table}
  \caption{Multi-wavelength broadband photometry of CEN\,34}
  \label{tbl1}
  \centering
  \begin{tabular}{c c c c c c}
    \hline\hline
    Band & $\lambda_{\mathrm{eff}}$ & mag & flux & Ref.\\
    &  $[$$\mu$m$]$ & & (mJy) & \\
    \hline
    B & 0.44 & 22.8$\pm$0.3  & 3.13$\pm$0.86$\times10^{-3}$ & 1  \\
    V & 0.55 & 18.82$\pm$0.05 & 0.114$\pm$0.005 & 1 \\
    Johnson\_R & 0.7 &   15.93$\pm0.05$     & 1.25$\pm0.06$  & 3        \\
    DENIS\_I & 0.79  & 14.482$\pm$0.04 &  5.5$\pm$0.2 & 2\\
    Johnson\_I & 0.9  &     13.38$\pm$0.05         & 12.0$\pm$0.5  & 3  \\
    2MASS\_J  & 1.23  & 10.402$\pm$0.024 & 110.0$\pm$2.4 & 4\\
    2MASS\_H  & 1.66  & 8.666$\pm$0.031  & 349.9$\pm$9.9 & 4\\
    2MASS\_K  & 2.16  & 7.788$\pm$0.023  & 511.4$\pm$10.8 & 4\\
    IRAC1 & 3.6 & 6.92$\pm$0.057 & 479$\pm$25 & 5\\
    L\arcmin    & 3.78 & 6.77  & 495 & 1\\
    IRAC2 & 4.5 & 6.425$\pm$0.043 & 483$\pm$19 & 5 \\
    IRAC3 & 5.8 & 5.834$\pm$0.034 & 541$\pm$17 & 5\\
    IRAC4 & 8.0 & 5.267$\pm$0.14 & 493$\pm$64 & 5\\
    N     & 10.5 &     & 2500$\pm$730  & 3 \\
    Q     & 18.6 &     & 5050$\pm$1395 & 3 \\
    \hline
  \end{tabular}
  \tablebib{(1)~\citet{2008Hoffmeister}; (2)~\citet{1997Msngr..87...27E}; (3)~\citet{1985A&A...146..175C}; (4) \citet{2006AJ....131.1163S}; (5)
    \citet{2003PASP..115..953B}.}
\end{table}

\subsection{Extinction and Luminosity\label{S:Exti}}
\begin{figure}
  \centering
  \includegraphics[width=0.48\textwidth]{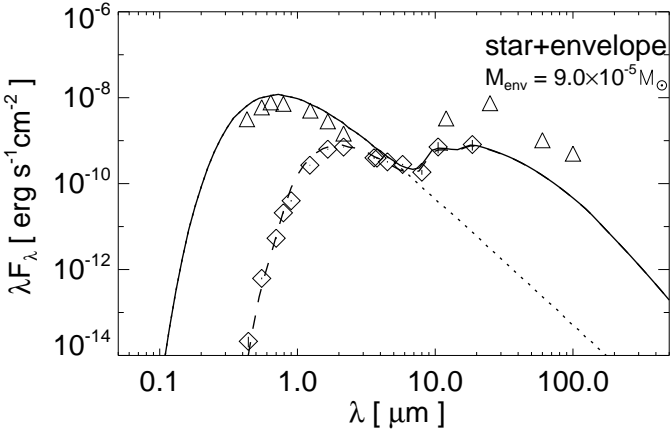}
  \caption{Observed SED of CEN\,34 generated from the photometric data listed in Table.~\ref{tbl1}~(diamonds), together with the best-fitting model SED (dashed line) based on a star+envelope system of $M_{env}=9\times10^{-5}M_{\sun}$. After dereddening~(A$_\mathrm{v}=11$~mag), the extinction free model SED~(solid line) can be compared to the derenddened fluxes of the reference post-AGB star IRAS~23304+6147~\citep[triangles;][]{2002A&A...385..884F}. In addition, the best-fitting stellar blackbody~($T_\mathrm{eff}=5260\,\mathrm{K}$ )is shown as a dotted line.}
  \label{Fig:Se}%
\end{figure}

In Table.~\ref{tbl1}, we collected the broadband photometric results for CEN\,34 available from both the literature and the catalogues of recent infrared surveys (DENIS, 2MASS and \textit{Spitzer} GLIMPSE) to construct the SED of CEN\,34 (see Fig.~\ref{Fig:Se}). The SED shows moderate infrared excess, most likely due to the presence of circumstellar material in an envelope and/or a disk. If CEN\,34 is an evolved star the circumstellar envelope is a more plausible assumption than a circumstellar disk \citep{2002A&A...385..884F}.

We use the simulated annealing approach described in \citet{2012A&A...546A...7L} to model the observed SED and to estimate the mass of circumstellar material. We explore the scenario of a spherical envelope that is commonly adopted to mimic the dust shell surrounding a post-AGB star. The corresponding dust density distribution is assumed to be a power law $\rho_{env}\propto r^{-\alpha}$ with $\alpha = 2$. Our best model fit is shown in Fig.~\ref{Fig:Se}. The observed SED~(diamonds in Fig.~\ref{Fig:Se}) at $\lambda{\lesssim}\,4\,\mu$m, i.e. dominated by the stellar radiation~(dotted line in Fig.~\ref{Fig:Se}), is reproduced by a star with $T_{\mathrm{eff}} = 5260~\mathrm{K}$ reddened by $A_V = 11$\,mag. The SED longward of $4\mu$m can be reproduced by a spherical envelope of $9 \times 10^{-5}\,M_{\sun}$, yielding an optical depth $\tau_v= 0.29$ for the circumstellar material along the line-of-sight. This corresponds to $A_V = 0.3$\,mag and is only a small fraction of the total extinction of $A_V = 11$\,mag mentioned above. Considering that \citet{2008ApJ...686..310H} derived a foreground extinction of 2\,mag toward M\,17 the largest fraction of extinction originates inside the \ion{H}{ii} region and behind it, in case CEN\,34 is not a cluster member. Therefore, the distance of CEN\,34 is at least greater than 2~kpc, the distance of M\,17. With this lower limit of CEN\,34's distance, the lower limit of the luminosity is determined as 1600~$L_{\sun}$. On the other hand, the derived temperature and surface gravity also constrain the absolute magnitude of CEN\,34. Using the calibration from \citep{1981Ap&SS..80..353S} we obtain $M_V=-4$~mag which transforms into a photometric distance of 2.3~kpc. However, this distance estimate has a large error due to the uncertainty of $\log g$.

\begin{figure*}
  \centering
  \includegraphics[width=0.7\textwidth]{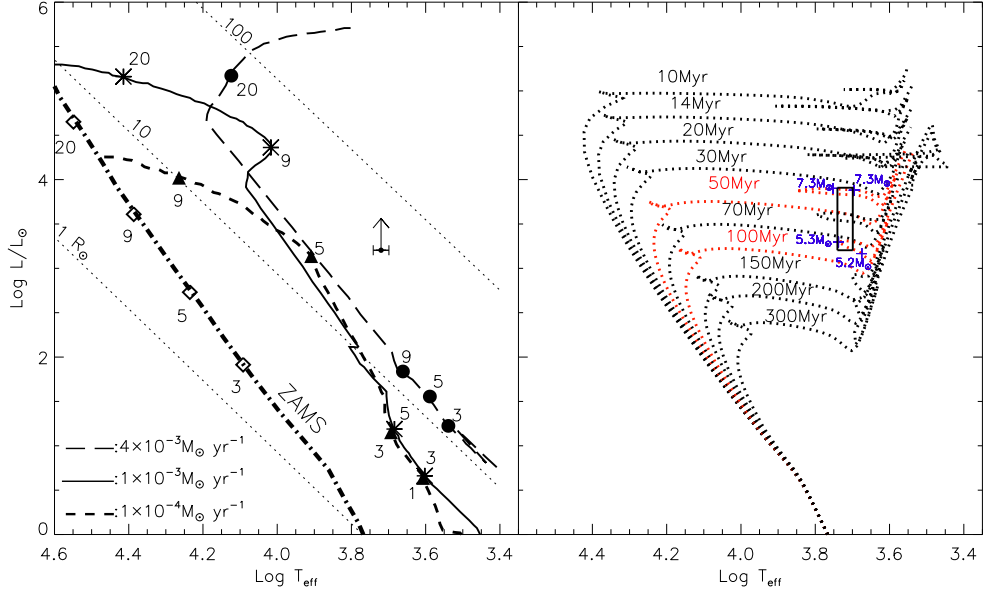}
  \caption{The position of CEN\,34 in the Hertzsprung-Russell diagram~(HRD) with well determined effective temperature~($5250\pm250~\mathrm{K}$) and loosely constrained luminosity~($1600~L_{\sun}\lesssim L_\ast \lesssim 8100~L_{\sun}$). Left panel: the position of CEN\,34 denoted by the full dot with error bar is compared with the model of massive star formation via disk accretion~\citep{2010ApJ...721..478H}. The luminosity of $1600~L_{\sun}$ adopted here is the lower limit~(denoted by the upward arrow) when the distance of CEN\,34 is assumed as 2\,kpc. Right panel: the position of CEN\,34~(the rectangle representing the uncertainties of $T_\mathrm{eff}$ and $L_\ast$) is placed against the isochrones of post-main-sequence evolution~\citep{2008A&A...482..883M}. The lower and upper isochrones~(roughly the 100~Myr and 50~Myr ones, respectively) that enclose the position of CEN\,34 are highlighted in red, and the corresponding masses are highlighted in blue.}
  \label{Fig:Hrd}%
\end{figure*}

\section{Discussion\label{S:Diss}}


In Fig.~\ref{Fig:Hrd} we outline the location of CEN\,34 in the Hertzsprung-Russell diagram~(HRD) based on effective temperature and luminosity ranges constrained in Sects.~\ref{S:Temp} and \ref{S:Exti}. Comparison is made to two distinct sets of theoretical evolutionary tracks and isochrones: while in the left panel, we compare to high-mass star formation models of \citet{2010ApJ...721..478H}, in the right panel we overplot post-main sequence isochrones of \citet{2008A&A...482..883M}. Obviously both high-mass YSOs and evolved stars are compatible with the observations of CEN\,34.

As demonstrated by the numerical simulations of \citet{2010ApJ...721..478H}, very young and accreting high-mass YSOs tend to have rather low $T_\mathrm{eff}$ and large radius, similar to those obtained for CEN\,34 in Sect.~\ref{S:Temp}. However, if we assume that CEN\,34 is associated with the M\,17 region and located at a distance of 2~kpc, its $T_\mathrm{eff}\approx5250$~K and luminosity $L_{\ast}\gtrsim1600~L_{\sun}$ place CEN\,34 in an area of the HRD where the numerical models suggest accretion rates in excess of $4\times10^{-3}M_{\sun}/$yr. Nevertheless, accretion rates may vary strongly with the evolution of high-mass YSOs. As a consequence, CEN\,34 needs not maintain such a high accretion rate throughout its entire formation process. Such extremely high accretion rates are usually expected to correlate with the presence of a very massive ($\sim1M_{\sun}$) envelope/disk surrounding the central high-mass YSO, which is obviously in contradiction to the amount of circumstellar material actually estimated for CEN\,34 ($\sim10^{-4}M_{\sun}$; see Sect.~\ref{S:Exti}). This discrepancy could be circumvented by a scenario in which the high-mass YSO has already accumulated most of its circumstellar material in a very short time span and, hence, the circumstellar material becomes optically thin. We speculate that CEN\,34 might be in a phase of gravitational contraction, during which it will not gain more material from its surroundings and will move to the left in HRD.

To solve the $V_{LSR}$ inconsistency between CEN\,34 and M\,17, one might argue that CEN\,34 has been ejected from M\,17. Recently \citet{2013MNRAS.430L..20G} reported that even O-type stars can be ejected from a massive cluster because of a dynamical few-body encounter in the cluster's core. If we propose that CEN\,34 is dynamically kicked out from the cluster center, then the time that CEN\,34 needs to move to the current position relative to the cluster center can be used to test our speculation. The distance from CEN\,34 to the cluster center is about 1~pc, which is the relative distance projected to the sky plane. Assuming 1~pc to the relative distance perpendicular to the sky plane, the $V_{LSR}$ difference of $\sim80$~km/s yields a time of $\sim1.2\times10^4$~yrs needed for CEN\,34 to move to the current position. This traveling time is comparable with the age of CEN\,34 if it had accumulated the bulk of its mass with accretion rate $>4\times10^{-3}M_{\sun}/$yr within a period of $\sim10^3$ yrs. A byproduct of this ejection scenario is to place CEN\,34 in an isolated environment where feeding through its parental cloud cannot be continued.


Alternatively, besides the fact that the spectrum of CEN\,34 shows characteristics very similar to those of post-AGB stars, which typically are late-type supergiants with spectral absorption features as discussed in Sect.~\ref{S:Spt}--\ref{S:Co} \citep{2007A&A...469..799S}, the overall shape of CEN\,34's SED is reminiscent of post-AGB stars. To foster the hypothesis of CEN\,34 being a post-AGB star, we compare its dereddened SED (solid lines in Fig.~\ref{Fig:Se}) to that of a reference post-AGB star of spectral type G2~Ia (\object{IRAS\,23304+6147}; triangles in Fig.~\ref{Fig:Se}) from \citet{2002A&A...385..884F}, located at a distance ($d\sim4.5$\,kpc). IRAS\,23304+6147 has mid-infrared excess slightly stronger than that of CEN\,34, most likely due to the presence of a larger amount of circumstellar material. In view of the similar spectral types and luminosity classes of CEN\,34 and IRAS\,23304+6147, their comparable optical and NIR fluxes imply a comparable distance too. Because the optical and NIR fluxes of CEN\,34 are still a bit higher than those of IRAS\,23304+6147, the distance of IRAS\,23304+6147 (i.e., 4.5 kpc) serves as upper limit for CEN\,34's distance. Considering the most likely range of CEN\,34's distance (2~kpc~$\lesssim$~d~$\lesssim$~4.5~kpc), the corresponding luminosity range is easily derived ($1600~L_{\sun}\lesssim L_\ast \lesssim 8100~L_{\sun}$). Comparing this luminosity range and the effective temperature to post-main sequence isochrones of \citet{2008A&A...482..883M} (see the right panel of Fig.~\ref{Fig:Hrd}), we constrain the ranges of stellar mass and age to 5\,--\,7~$M_{\sun}$ and 100\,--\,50~Myr, respectively. This would suggest that CEN\,34 is an evolved intermediate-mass star. While post-AGB stars typically yield prominent far-infrared emissions \citep[e.g.,][]{2006A&A...458..173S,2007A&A...469..799S}, far-infrared flux measurements of CEN\,34 (e.g., from IRAS or AKARI) are highly contaminated by the strong dust continuum of the M\,17 S-bar. For this reason, the 60\,$\mu$m and 100\,$\mu$m fluxes listed for CEN\,34 in the IRAS Point Source Reject Catalog (released in 1989) are unreliable, because of the large IRAS beam size ($\sim1$ arcmin at 60\,$\mu$m); although AKARI observed with higher spatial resolution and better sensitivities than IRAS, unfortunately the AKARI/FIS All-Sky Survey Point Source Catalogue~\citep{2010yCat.2298....0Y} does not return an entry matching the position of CEN\,34.

\section{Summary and Conclusions\label{S:CONCL}}

CEN\,34\,--\,originally considered a high-mass YSO candidate in the M\,17 star forming region\,--\,is re-classified as an early-/mid-G supergiant on the basis of the new X-Shooter spectrum. The SED of CEN\,34 resembles that of either a high-mass YSO or an evolved intermediate-mass star (likely a post-AGB star) with a visual extinction $A_V = 11$\,mag that it is dominated by interstellar extinction. If CEN\,34 is a high-mass YSO, its derive properties ($T_\mathrm{eff}$, $\log g$, and $V_{LSR}$) can be explained by a scenario where the star has been ejected from the M\,17 cluster. In this case CEN\,34 had accumulated the bulk of its mass during the first $\sim10^3$~years and it is currently gravitationally contracting with only little circumstellar material left. If, on the other hand, CEN\,34 is an evolved star of intermediate mass, its luminosity is in the range $1600\,L_{\sun} \lesssim L_\ast \lesssim 8100\,L_{\sun}$, corresponding to a distance range of 2\,kpc~$\lesssim d \lesssim 4.5$\,kpc. Comparison to the isochrones of post-main sequence evolution then suggests that CEN\,34 has a mass of $5 -7\,M_{\sun}$ and an age of $50 - 100$\,Myr, which are reminiscent of a post-AGB star. 

In conclusion, the ambiguities concerning infrared excess and spectral types between high-mass YSOs and evolved intermediate mass stars prevent a unique classification of CEN\,34. The arguments of interpreting CEN\,34 as a high-mass YSO are not as straightforward as in the case of a post-AGB star, and they are based on a number of assumptions. For the time being, we prefer the interpretation of a post-AGB star and hope that a future more accurate distance estimate will settle the issue.

\begin{acknowledgements}
We acknowledge the anonymous referee whose comments helped to clarify the paper at several points. The ESO Paranal staff is acknowledged for obtaining the X-Shooter spectrum of CEN\,34 in Service Mode. M.F. acknowledges the support by NSFC through grants 11203081. This publication makes use of data products from the Two Micron All Sky Survey, which is a joint project of the University of Massachusetts and the Infrared Processing and Analysis Center/California Institute of Technology, funded by the National Aeronautics and Space Administration and the National Science Foundation. This research has made use of the SIMBAD database, operated at CDS, Strasbourg, France. This research has made use of the NASA/IPAC Infrared Science Archive (IRSA) which is operated by the Jet Propulsion Laboratory, California Institute of Technology, under contract with the National Aeronautics and Space Administration. This research has made use of NASA's Astrophysics Data System Bibliographic Services.
\end{acknowledgements}

\bibliographystyle{aa} \bibliography{myrefs}

\begin{thebibliography}{46}
\expandafter\ifx\csname natexlab\endcsname\relax\def\natexlab#1{#1}\fi

\bibitem[{{Andrillat} {et~al.}(1995){Andrillat}, {Jaschek}, \&
  {Jaschek}}]{1995A&AS..112..475A}
{Andrillat}, Y., {Jaschek}, C., \& {Jaschek}, M. 1995, \aaps, 112, 475

\bibitem[{{Benjamin} {et~al.}(2003){Benjamin}, {Churchwell}, {Babler}, {Bania},
  {Clemens}, {Cohen}, {Dickey}, {Indebetouw}, {Jackson}, {Kobulnicky},
  {Lazarian}, {Marston}, {Mathis}, {Meade}, {Seager}, {Stolovy}, {Watson},
  {Whitney}, {Wolff}, \& {Wolfire}}]{2003PASP..115..953B}
{Benjamin}, R.~A., {Churchwell}, E., {Babler}, B.~L., {et~al.} 2003, \pasp,
  115, 953

\bibitem[{{Carquillat} {et~al.}(1997){Carquillat}, {Jaschek}, {Jaschek}, \&
  {Ginestet}}]{1997A&AS..123....5C}
{Carquillat}, M.~J., {Jaschek}, C., {Jaschek}, M., \& {Ginestet}, N. 1997,
  \aaps, 123, 5

\bibitem[{{Casali} \& {Eiroa}(1996)}]{1996A&A...306..427C}
{Casali}, M.~M. \& {Eiroa}, C. 1996, \aap, 306, 427

\bibitem[{{Chen} {et~al.}(2012){Chen}, {Jiang}, {Wang}, {Chini}, {Tamura},
  {Nagayama}, {Nagata}, \& {Nakajima}}]{2012PASJ...64...110C}
{Chen}, Z., {Jiang}, Z., {Wang}, Y., {et~al.} 2012, \pasj, 64, 110

\bibitem[{{Chini} {et~al.}(2004){Chini}, {Hoffmeister}, {Kimeswenger},
  {Nielbock}, {N{\"u}rnberger}, {Schmidtobreick}, \&
  {Sterzik}}]{2004Natur.429..155C}
{Chini}, R., {Hoffmeister}, V., {Kimeswenger}, S., {et~al.} 2004, \nat, 429,
  155

\bibitem[{{Chini} \& {Kr{\"u}gel}(1985)}]{1985A&A...146..175C}
{Chini}, R. \& {Kr{\"u}gel}, E. 1985, \aap, 146, 175

\bibitem[{{Dame} \& {Thaddeus}(2011)}]{2011ApJ...734L..24D}
{Dame}, T.~M. \& {Thaddeus}, P. 2011, \apjl, 734, L24

\bibitem[{{De Medeiros} {et~al.}(2002){De Medeiros}, {Udry}, {Burki}, \&
  {Mayor}}]{2002A&A...395...97D}
{De Medeiros}, J.~R., {Udry}, S., {Burki}, G., \& {Mayor}, M. 2002, \aap, 395,
  97

\bibitem[{{D'Odorico} {et~al.}(2006){D'Odorico}, {Dekker}, {Mazzoleni},
  {Vernet}, {Guinouard}, {Groot}, {Hammer}, {Rasmussen}, {Kaper}, {Navarro},
  {Pallavicini}, {Peroux}, \& {Zerbi}}]{2006SPIE.6269E..98D}
{D'Odorico}, S., {Dekker}, H., {Mazzoleni}, R., {et~al.} 2006, in Society of
  Photo-Optical Instrumentation Engineers (SPIE) Conference Series, Vol. 6269,
  Society of Photo-Optical Instrumentation Engineers (SPIE) Conference Series

\bibitem[{{Epchtein} {et~al.}(1997){Epchtein}, {de Batz}, {Capoani},
  {Chevallier}, {Copet}, {Fouqu{\'e}}, {Lacombe}, {Le Bertre}, {Pau}, {Rouan},
  {Ruphy}, {Simon}, {Tiph{\`e}ne}, {Burton}, {Bertin}, {Deul}, {Habing},
  {Borsenberger}, {Dennefeld}, {Guglielmo}, {Loup}, {Mamon}, {Ng}, {Omont},
  {Provost}, {Renault}, {Tanguy}, {Kimeswenger}, {Kienel}, {Garzon}, {Persi},
  {Ferrari-Toniolo}, {Robin}, {Paturel}, {Vauglin}, {Forveille}, {Delfosse},
  {Hron}, {Schultheis}, {Appenzeller}, {Wagner}, {Balazs}, {Holl},
  {L{\'e}pine}, {Boscolo}, {Picazzio}, {Duc}, \&
  {Mennessier}}]{1997Msngr..87...27E}
{Epchtein}, N., {de Batz}, B., {Capoani}, L., {et~al.} 1997, The Messenger, 87,
  27

\bibitem[{{Figer} {et~al.}(2003){Figer}, {Gilmore}, {Kim}, {Morris}, {Becklin},
  {McLean}, {Gilbert}, {Graham}, {Larkin}, {Levenson}, \&
  {Teplitz}}]{2003ApJ...599.1139F}
{Figer}, D.~F., {Gilmore}, D., {Kim}, S.~S., {et~al.} 2003, \apj, 599, 1139

\bibitem[{{Fujii} {et~al.}(2002){Fujii}, {Nakada}, \&
  {Parthasarathy}}]{2002A&A...385..884F}
{Fujii}, T., {Nakada}, Y., \& {Parthasarathy}, M. 2002, \aap, 385, 884

\bibitem[{{Goldoni}(2011)}]{2011AN....332..227G}
{Goldoni}, P. 2011, Astronomische Nachrichten, 332, 227

\bibitem[{{Gonz{\'a}lez-Fern{\'a}ndez} \&
  {Negueruela}(2012)}]{2012A&A...539A.100G}
{Gonz{\'a}lez-Fern{\'a}ndez}, C. \& {Negueruela}, I. 2012, \aap, 539, A100

\bibitem[{{Gray} \& {Corbally}(2009)}]{2009ssc..book.....G}
{Gray}, R.~O. \& {Corbally}, J.~C. 2009, {Stellar Spectral Classification}

\bibitem[{{Gvaramadze} {et~al.}(2013){Gvaramadze}, {Kniazev}, {Chen{\'e}}, \&
  {Schnurr}}]{2013MNRAS.430L..20G}
{Gvaramadze}, V.~V., {Kniazev}, A.~Y., {Chen{\'e}}, A.-N., \& {Schnurr}, O.
  2013, \mnras, 430, L20

\bibitem[{{Hanson} {et~al.}(1997){Hanson}, {Howarth}, \&
  {Conti}}]{1997ApJ...489..698H}
{Hanson}, M.~M., {Howarth}, I.~D., \& {Conti}, P.~S. 1997, \apj, 489, 698

\bibitem[{{Herbig}(1977)}]{1977ApJ...214..747H}
{Herbig}, G.~H. 1977, \apj, 214, 747

\bibitem[{{Hoffmeister}(2008)}]{2008Hoffmeister}
{Hoffmeister}, V.~H. 2008, PhD thesis, Ruhr-Universit{\"a}t Bochum

\bibitem[{{Hoffmeister} {et~al.}(2006){Hoffmeister}, {Chini}, {Scheyda},
  {N{\"u}rnberger}, {Vogt}, \& {Nielbock}}]{2006A&A...457L..29H}
{Hoffmeister}, V.~H., {Chini}, R., {Scheyda}, C.~M., {et~al.} 2006, \aap, 457,
  L29

\bibitem[{{Hoffmeister} {et~al.}(2008){Hoffmeister}, {Chini}, {Scheyda},
  {Schulze}, {Watermann}, {N{\"u}rnberger}, \& {Vogt}}]{2008ApJ...686..310H}
{Hoffmeister}, V.~H., {Chini}, R., {Scheyda}, C.~M., {et~al.} 2008, \apj, 686,
  310

\bibitem[{{Hosokawa} {et~al.}(2010){Hosokawa}, {Yorke}, \&
  {Omukai}}]{2010ApJ...721..478H}
{Hosokawa}, T., {Yorke}, H.~W., \& {Omukai}, K. 2010, \apj, 721, 478

\bibitem[{{Jiang} {et~al.}(2002){Jiang}, {Yao}, {Yang}, {Ando}, {Kato},
  {Kawai}, {Kurita}, {Nagata}, {Nagayama}, {Nakajima}, {Nagashima}, {Sato},
  {Tamura}, {Nakaya}, \& {Sugitani}}]{2002ApJ...577..245J}
{Jiang}, Z., {Yao}, Y., {Yang}, J., {et~al.} 2002, \apj, 577, 245

\bibitem[{{Liu} {et~al.}(2012){Liu}, {Madlener}, {Wolf}, {Wang}, \&
  {Ruge}}]{2012A&A...546A...7L}
{Liu}, Y., {Madlener}, D., {Wolf}, S., {Wang}, H., \& {Ruge}, J.~P. 2012, \aap,
  546, A7

\bibitem[{{Marigo} {et~al.}(2008){Marigo}, {Girardi}, {Bressan}, {Groenewegen},
  {Silva}, \& {Granato}}]{2008A&A...482..883M}
{Marigo}, P., {Girardi}, L., {Bressan}, A., {et~al.} 2008, \aap, 482, 883

\bibitem[{{Munari} {et~al.}(2005){Munari}, {Sordo}, {Castelli}, \&
  {Zwitter}}]{2005A&A...442.1127M}
{Munari}, U., {Sordo}, R., {Castelli}, F., \& {Zwitter}, T. 2005, \aap, 442,
  1127

\bibitem[{{Munari} \& {Tomasella}(1999)}]{1999A&AS..137..521M}
{Munari}, U. \& {Tomasella}, L. 1999, \aaps, 137, 521

\bibitem[{{Nielbock} {et~al.}(2007){Nielbock}, {Chini}, {Hoffmeister},
  {Scheyda}, {Steinacker}, {N{\"u}rnberger}, \&
  {Siebenmorgen}}]{2007ApJ...656L..81N}
{Nielbock}, M., {Chini}, R., {Hoffmeister}, V.~H., {et~al.} 2007, \apjl, 656,
  L81

\bibitem[{{Nielbock} {et~al.}(2001){Nielbock}, {Chini}, {J{\"u}tte}, \&
  {Manthey}}]{2001A&A...377..273N}
{Nielbock}, M., {Chini}, R., {J{\"u}tte}, M., \& {Manthey}, E. 2001, \aap, 377,
  273

\bibitem[{{N{\"u}rnberger} {et~al.}(2007){N{\"u}rnberger}, {Chini},
  {Eisenhauer}, {Kissler-Patig}, {Modigliani}, {Siebenmorgen}, {Sterzik}, \&
  {Szeifert}}]{2007A&A...465..931N}
{N{\"u}rnberger}, D.~E.~A., {Chini}, R., {Eisenhauer}, F., {et~al.} 2007, \aap,
  465, 931

\bibitem[{{Ochsendorf} {et~al.}(2011){Ochsendorf}, {Ellerbroek}, {Chini},
  {Hartoog}, {Hoffmeister}, {Waters}, \& {Kaper}}]{2011A&A...536L...1O}
{Ochsendorf}, B.~B., {Ellerbroek}, L.~E., {Chini}, R., {et~al.} 2011, \aap,
  536, L1

\bibitem[{{Povich} {et~al.}(2009){Povich}, {Churchwell}, {Bieging}, {Kang},
  {Whitney}, {Brogan}, {Kulesa}, {Cohen}, {Babler}, {Indebetouw}, {Meade}, \&
  {Robitaille}}]{2009ApJ...696.1278P}
{Povich}, M.~S., {Churchwell}, E., {Bieging}, J.~H., {et~al.} 2009, \apj, 696,
  1278

\bibitem[{{Reader} {et~al.}(1980){Reader}, {Corliss}, {Wiese}, \&
  {Martin}}]{1980wtpa.book.....R}
{Reader}, J., {Corliss}, C.~H., {Wiese}, W.~L., \& {Martin}, G.~A. 1980,
  {Wavelengths and transition probabilities for atoms and atomic ions: Part 1.
  Wavelengths, part 2. Transition probabilities}

\bibitem[{{Reid} {et~al.}(2009){Reid}, {Menten}, {Zheng}, {Brunthaler},
  {Moscadelli}, {Xu}, {Zhang}, {Sato}, {Honma}, {Hirota}, {Hachisuka}, {Choi},
  {Moellenbrock}, \& {Bartkiewicz}}]{2009ApJ...700..137R}
{Reid}, M.~J., {Menten}, K.~M., {Zheng}, X.~W., {et~al.} 2009, \apj, 700, 137

\bibitem[{{Skrutskie} {et~al.}(2006){Skrutskie}, {Cutri}, {Stiening},
  {Weinberg}, {Schneider}, {Carpenter}, {Beichman}, {Capps}, {Chester},
  {Elias}, {Huchra}, {Liebert}, {Lonsdale}, {Monet}, {Price}, {Seitzer},
  {Jarrett}, {Kirkpatrick}, {Gizis}, {Howard}, {Evans}, {Fowler}, {Fullmer},
  {Hurt}, {Light}, {Kopan}, {Marsh}, {McCallon}, {Tam}, {Van Dyk}, \&
  {Wheelock}}]{2006AJ....131.1163S}
{Skrutskie}, M.~F., {Cutri}, R.~M., {Stiening}, R., {et~al.} 2006, \aj, 131,
  1163

\bibitem[{{Straizys} \& {Kuriliene}(1981)}]{1981Ap&SS..80..353S}
{Straizys}, V. \& {Kuriliene}, G. 1981, \apss, 80, 353

\bibitem[{{Su{\'a}rez} {et~al.}(2006){Su{\'a}rez}, {Garc{\'{\i}}a-Lario},
  {Manchado}, {Manteiga}, {Ulla}, \& {Pottasch}}]{2006A&A...458..173S}
{Su{\'a}rez}, O., {Garc{\'{\i}}a-Lario}, P., {Manchado}, A., {et~al.} 2006,
  \aap, 458, 173

\bibitem[{{Szczerba} {et~al.}(2007){Szczerba}, {Si{\'o}dmiak}, {Stasi{\'n}ska},
  \& {Borkowski}}]{2007A&A...469..799S}
{Szczerba}, R., {Si{\'o}dmiak}, N., {Stasi{\'n}ska}, G., \& {Borkowski}, J.
  2007, \aap, 469, 799

\bibitem[{{Valdes} {et~al.}(2004){Valdes}, {Gupta}, {Rose}, {Singh}, \&
  {Bell}}]{2004ApJS..152..251V}
{Valdes}, F., {Gupta}, R., {Rose}, J.~A., {Singh}, H.~P., \& {Bell}, D.~J.
  2004, \apjs, 152, 251

\bibitem[{{Vernet} {et~al.}(2011){Vernet}, {Dekker}, {D'Odorico}, {Kaper},
  {Kjaergaard}, {Hammer}, {Randich}, {Zerbi}, {Groot}, {Hjorth}, {Guinouard},
  {Navarro}, {Adolfse}, {Albers}, {Amans}, {Andersen}, {Andersen}, {Binetruy},
  {Bristow}, {Castillo}, {Chemla}, {Christensen}, {Conconi}, {Conzelmann},
  {Dam}, {de Caprio}, {de Ugarte Postigo}, {Delabre}, {di Marcantonio},
  {Downing}, {Elswijk}, {Finger}, {Fischer}, {Flores}, {Fran{\c c}ois},
  {Goldoni}, {Guglielmi}, {Haigron}, {Hanenburg}, {Hendriks}, {Horrobin},
  {Horville}, {Jessen}, {Kerber}, {Kern}, {Kiekebusch}, {Kleszcz}, {Klougart},
  {Kragt}, {Larsen}, {Lizon}, {Lucuix}, {Mainieri}, {Manuputy}, {Martayan},
  {Mason}, {Mazzoleni}, {Michaelsen}, {Modigliani}, {Moehler}, {M{\o}ller},
  {Norup S{\o}rensen}, {N{\o}rregaard}, {P{\'e}roux}, {Patat}, {Pena}, {Pragt},
  {Reinero}, {Rigal}, {Riva}, {Roelfsema}, {Royer}, {Sacco}, {Santin},
  {Schoenmaker}, {Spano}, {Sweers}, {Ter Horst}, {Tintori}, {Tromp}, {van
  Dael}, {van der Vliet}, {Venema}, {Vidali}, {Vinther}, {Vola}, {Winters},
  {Wistisen}, {Wulterkens}, \& {Zacchei}}]{2011A&A...536A.105V}
{Vernet}, J., {Dekker}, H., {D'Odorico}, S., {et~al.} 2011, \aap, 536, A105

\bibitem[{{Wallace} \& {Hinkle}(1997)}]{1997ApJS..111..445W}
{Wallace}, L. \& {Hinkle}, K. 1997, \apjs, 111, 445

\bibitem[{{Wilson} {et~al.}(2003){Wilson}, {Hanson}, \&
  {Muders}}]{2003ApJ...590..895W}
{Wilson}, T.~L., {Hanson}, M.~M., \& {Muders}, D. 2003, \apj, 590, 895

\bibitem[{{Xu} {et~al.}(2011){Xu}, {Moscadelli}, {Reid}, {Menten}, {Zhang},
  {Zheng}, \& {Brunthaler}}]{2011ApJ...733...25X}
{Xu}, Y., {Moscadelli}, L., {Reid}, M.~J., {et~al.} 2011, \apj, 733, 25

\bibitem[{{Yamamura} {et~al.}(2010){Yamamura}, {Makiuti}, {Ikeda}, {Fukuda},
  {Oyabu}, {Koga}, \& {White}}]{2010yCat.2298....0Y}
{Yamamura}, I., {Makiuti}, S., {Ikeda}, N., {et~al.} 2010, VizieR Online Data
  Catalog, 2298, 0

\bibitem[{{Zinnecker} \& {Yorke}(2007)}]{2007ARA&A..45..481Z}
{Zinnecker}, H. \& {Yorke}, H.~W. 2007, \araa, 45, 481

\end{thebibliography}

\end{document}